\documentclass{aastex}          
\usepackage{spr-astr-addons}    

\begin{document}
%
\title{Evidence against non-gravitational acceleration of 1I/2017 U1
`Oumuamua}

\shorttitle{Non-gravitational acceleration of `Oumuamua?}
\shortauthors{Katz}

\author{J. I. Katz\altaffilmark{}} 
\affil{Department of Physics and McDonnell Center for the Space Sciences\\
Washington University, St. Louis, Mo. 63130}
\email{katz@wuphys.wustl.edu} 
\date{\today}

\begin{abstract}
\citet{M18} reported that a seven-parameter fit to the orbit of
1I/2017 U1 `Oumuamua indicated a non-gravitational acceleration
in the anti-Solar direction, and attributed it to recoil from
comet-like outgassing.  The implied gas to dust ratio is at least
100 times greater than that of known Solar System comets.  The
reported collapse of the scatter of nearly contemporaneous
coordinate residuals upon inclusion of the non-gravitational term in
the orbital fits is difficult to understand.  There are grounds for
skepticism.
\end{abstract}

\newpage
\citet{M17} discovered the interstellar object 1I/2017 U1 (`Oumuamua) and
estimated an upper limit of $1.7 \times 10^{-3}$ { kg/s} on its rate of
outflow of dust, describing it as an asteroid rather than a comet.  However,
\citet{M18} added a possible non-gravitational force, directed away from the
Sun with an inverse square law dependence on distance, to the orbital fit and
found a significant amplitude, effectively multiplying the Solar
gravitational acceleration by a factor of 0.99917.  Such a force might be
expected if `Oumuamua is outgassing under the influence of Solar radiative
heating, although a quantitative inverse square law might not be expected
because of the complexity of heat transfer in a transiently heated body.

From this result \citet{M18} inferred an outgassing rate of 3.6 kg/s,
implying a ratio of dust to gas in the outflow $< 0.5 \times 10^{-3}$.  This
should be compared to the dust to gas ratios of Solar System comets that are
0.1--1 \citep{SAH92,SSH96}.

\cite{M18} also set an upper bound of 1 kg on the dust present within a
cylinder of radius $r = 750$ km ($2.5^{\prime\prime}$) projected separation.
Using their outgassing rate of ${\dot M}_{gas} = 3.6$ kg/s and { assuming
roughly spherical symmetry of outgassing and} a plausible
ejection speed $v = 300$ m/s leads to a mass of gas in this cylinder
\begin{equation}
M_{gas} = {{\dot M} \over 4 \pi} \int\,d\Omega {r \over
v \sin\theta} = {\pi \over 2}{{\dot M} r \over v} = 1.4 \times
10^4\ \text{kg},
\end{equation}
where $\theta$ is the angle from the direction to the observer.  This
would imply a dust to gas ratio $< 10^{-4}$, even less than that inferred
by comparing their outgassing rate to the bound on the dust outflow rate
of \citet{M17}.

The claimed outgassing rate would imply that `Oumuamua consists of
extraordinarily clean ices, unlike any known Solar System { cometary
body}.  This is a reason for skepticism of the reported non-gravitational
acceleration.  { \cite{M18} have suggested that the gas flow might
contain comparatively large solid particles that are inefficient scatterers,
thus maintaining a solid to gas ratio comparable to that of Solar System
cometary bodies while producing very little visible coma.  This cannot be
excluded empirically, but the phenomenological difference is in the ratio
of fine dust to gas; even if these larger particles are present, `Oumuamua
would still differ from Solar System bodies in its ratio of fine dust to
gas.  The possibility of such larger particles is constrained by infrared
observations \citep{T18}.}

An additional reason for skepticism is that Fig.~2 of \citet{M18} shows
a scatter in the residuals to their 6-element (purely gravitational) orbital
fit in observations within a single night or on consecutive nights of $\sim 3$
times their formal uncertainty.  Adding the seventh fitting parameter, the
magnitude of the non-gravitational force, collapses this scatter by at least
a factor of three.  This would not be expected, even were the
non-gravitational forces correct, because they are a smoothly varying
function of time.  Including it in the fit (and optimizing over all seven
parameters) would not be expected to collapse the scatter of nearly
contemporaneous observations---the fitted orbit, would, at best, accurately
agree with the mean of such observations, but would not be expected to
reduce their scatter.

The validity of the fitting procedure can be tested by numerical experiment:
Add random noise with dispersion three times the formal uncertainty to
the six observations obtained on UTC 2017-10-19.  Then perform the six and
seven parameter fits to the entire dataset, including these test ``data''.
If the fitting procedure is valid, the scatter of the contemporaneous data
about the best fit solutions should not collapse in the seven parameter fit
because the test ``data'' would not be consistent with the physical model
(gravity plus outgassing).  If the scatter collapses, the fitting procedure
requires further investigation.

{ Nongravitational acceleration would be explicable if the column density
of `Oumuamua is $\sim 0.1\text{--}1$ g/cm$^2$.  \citet{BL18} have suggested
that it is a thin artificial structure, although the inferred column density
would be orders of magnitude greater than that of engineered ``Solar
Sails''.  \citet{M19} suggested it is a fractal aggregate of density $\sim
10^{-5}$ g/cm$^3$, although such an object would be extremely fragile and
would have much lower density than the microscopic fractal aggregates
observed in the Solar System.}

The arguments presented here are independent of that of \cite{R18}
but lead to the same conclusion: It is unlikely that `Oumuamua has a
non-gravitational acceleration as large as that reported by \citet{M18}.

\acknowledgments
I thank S. Bialy, S. Kenyon\break and A. Loeb for useful discussions.  This
is a\break post-peer-review, pre-copyedit version of an article\break
published in Astrophysics and Space Sciences.  The\break final authenticated
version is available online at\break
\url{http://dx.doi.org/10.1007/s10509-019-3542-z}.


%

\begin{thebibliography}{9}
	\bibitem[\protect\citeauthoryear{Bialy \& Loeb}{2018}]{BL18} Bialy,
		S. \& Loeb, A. 2018 \apjl\ 868, L1.
	\bibitem[\protect\citeauthoryear{Meech {\it et al.\/}}{2017}]
		{M17} Meech, K. J. {\it et al.\/} Nature 552, 378 (2017).
	\bibitem[\protect\citeauthoryear{Micheli {\it et al.\/}}{2018}]
		{M18} Micheli, M. {\it et al.\/} Nature 559, 223 (2018).
	\bibitem[\protect\citeauthoryear{Moro-Mart\'{\i}n}{2019}]{M19}
		Moro-Mart\'{\i}n, A. \apjl\ 872, L32.
	\bibitem[\protect\citeauthoryear{Rafikov}{2018}]{R18}Rafikov, R. R.
		Ap. J. Lett. 867, L17 (2018).
	\bibitem[\protect\citeauthoryear{Sanzovo, Singh \& Huebner}{1996}]
		{SSH96} Sanzovo, G. C., Singh, P. D. \& Huebner, W. F.
		Astron. Ap. Suppl. 120, 301 (1996).
	\bibitem[\protect\citeauthoryear{Singh, de Almeida \& Huebner}
		{1992}]{SAH92} Singh, P. D., de Almeida, A. A. \& Huebner,
		W. F. Astron. J. 104, 848 (1992).
	\bibitem[\protect\citeauthoryear{Trilling {\it et al.\/}}{2018}]
		{T18} Trilling, D. E., Mommert, M., Hora, J. L. {\it et
		al.\/} 2018 \aj\ 156, 261.
\end{thebibliography}

%
		\section*{Compliance with Ethical Standards}
		The author has no potential conflicts of interest.

\end{document}